# On Income Inequality and Population Size

**Thitithep Sitthiyot**[*]
Director
Public Debt Policy Research Division
Policy and Planning Bureau
Public Debt Management Office
Ministry of Finance
Thailand
thitithep@mof.go.th

**Kanyarat Holasut**
Associate Professor
Department of Chemical Engineering
Faculty of Engineering
Khon Kaen University
Thailand
kanyarat@kku.ac.th

[*] Corresponding author and submitted as an independent author. The contents of this article do not reflect the authors' affiliations. The authors are grateful to Suradit Holasut for initiating the idea and for valuable suggestions. The authors also thank two anonymous referees for their useful comments and criticisms. All errors rest with the authors.

# ABSTRACT

The pursuit of having an appropriate level of income inequality should be viewed as one of the biggest challenges facing academic scholars as well as policy makers. Unfortunately, research on this issue is currently lacking. This study is the first to introduce the theoretical concept of targeted level of income inequality for a given size of population. By employing the World Bank's data on population size and Gini coefficient from sixty-nine countries in 2012, this study finds that the relationship between Gini coefficient and natural logarithm of population size is nonlinear in the form of a second degree polynomial function. The estimated results using regression analysis show that the majority of countries in the sample have Gini coefficients either too high or too low compared to their appropriate values. These findings could be used as a guideline for policy makers before designing and implementing public policies in order to achieve the targeted level of income inequality.

## 1. Introduction

It is widely agreed among academic scholars and practitioners that the most commonly used measurement of income inequality is Gini coefficient. While the theoretical value of Gini coefficient lies between zero and one, in practice, the minimum and maximum values the Gini coefficient could possibly attain are zero and (P-1)/P, where P is the number of population. This could be illustrated by using an example of a hypothetical country. If a country has only one population, then there is obviously no income inequality and the value of Gini coefficient would be zero. That is the minimum value of Gini coefficient this country could attain. However, as the number of population gets larger, say, 2, 3, 5, 8, …, or P, and only one person has all the income while others have none, a situation of perfect income inequality, the maximum value of Gini coefficient for this hypothetical country to attain would be 1/2, 2/3, 4/5, 7/8, …, or (P-1)/P, respectively.[1] In practice, the value of Gini coefficient, therefore, should be greater than zero but less than (P-1)/P.

The above example indicates that, theoretically, there should be an association between the degree of income inequality as measured by Gini coefficient and the size of population. This is consistent with Deltas (2003) who argues that the Gini coefficient of a small population would be smaller than that of a larger one generated by the same stochastic process. Equivalently, removing members of a population at random would tend to lower the estimated Gini coefficient of that population. Deltas also notes that, for any given level of intrinsic inequality, as expressed by income generating function, a reduction in the sample size would lead to a reduction in inequality as measured by the Gini coefficient. In addition, countries with small populations and less diverse economies tend to report small Gini coefficients whereas a much higher Gini coefficient are expected for countries with economically diverse large populations ("Gini Coefficient", 2016).

---

[1] This could simply be calculated geometrically by dividing the area between the 45-degree line and the Lorenz curve by 1/2.

While there are empirical research examining the relationship between income inequality and size of population or size of state as well as other economic, social, and political variables,[2] the issue of what an appropriate degree of income inequality as measured by the Gini coefficient should be for a country given a population size has yet to be explored by the existing literatures.[3] According to the income inequality and population data in 2012 compiled by the World Bank (2016a; 2016b), countries that have similar values of Gini coefficient could have very different population size. For example, Bhutan, with population of only 743,711, has the Gini coefficient of 0.387 while Thailand, with population of 67,164,130, has a slightly higher value of the Gini coefficient of 0.393. Does this imply that income inequality in Bhutan is not much different from that in Thailand? The same World Bank's data also show that countries that are similar in terms of population size could have very different level of income inequality. For example, Guinea, with population of 11,628,767, has Gini coefficient of 0.337 whereas Haiti, with a slightly lower population of 10,288,828, has almost twice the value of Gini coefficient at 0.608. Based on these observations, can we conclude that people of Guinea has more income equality than those of Haiti?

---

[2] For studies that focus on the issue of income inequality and size of population or size of state, please see Streeten (1993), Commonwealth Secretariat (2000), Bräutigam and Woolcock (2001), Alesina (2003), and Campante and Do (2007). For those that investigate the relationship between income inequality and other social, economic, and political factors, please see Phongpaichit (2016) and references therein.

[3] To the best of the authors' knowledge, as of this writing, the authors have found no study that investigates the appropriate level of income inequality as measured by Gini coefficient for a given size of population.

Given that there are various economic, social, and political factors that could have effects on income inequality as investigated by earlier research,[4] it is interesting to examine the linkage between the degree of income inequality as measured by Gini coefficient and the population size, and find out empirically an appropriate value of Gini coefficient given the size of population since no study has been conducted thus far. With an exception of two extreme cases of perfect income equality and perfect income inequality regardless of population size, this study hypothesizes that a country with small populations should have relatively lower Gini coefficient than a country with large populations due to the degree of economic, social, and political diversities as already reflected by the size of population. This study views that knowing the appropriate level of income inequality as measured by Gini coefficient could benefit policy makers as a starting point that can be used as a guideline prior to design and implement public policies to tackle the issue of income inequality or income equality.

This study is organized into five sections. Following the Introduction, Section 2 discusses the logic of the appropriate degree of income inequality. Section 3 explains research methodology and data employed in this study. Section 4 presents empirical findings and discusses the issue of targeted level of income inequality. Finally, Section 5 concludes and provides policy implications as well as suggestions for future research.

---

[4] See Streeten (1993), Commonwealth Secretariat (2000), Bräutigam and Woolcock (2001), Alesina (2003), Campante and Do (2007), and Phongpaichit (2016).

## 2. The Logic of Appropriate Degree of Income Inequality

High or extreme income inequality, theoretically, could cause economic, social, and political disruptions in many ways. Fuentes-Nieva and Galasso (2014) criticizes that extreme economic inequality is damaging to society for several reasons. It could have negative impacts on growth and poverty reduction. Extreme economic disparity is also worrying because of the pernicious impact that wealth concentrations could have on equal political representation. When wealth dominates public policymaking, the laws and regulations are bent to favor the rich and often to the detriment of the rest in the society. Equally alarming, public opinion could be shaped and election outcome could be affected by large-scale propaganda efforts through media the rich own or can control (Raza, 2016). According to Fuentes-Nieva and Galasso (2014), these could lead to the erosion of democratic governance, the pulling apart of social cohesion, and the vanishing of equal opportunities for all. Left unchecked, the adverse effects of high or extreme income inequality are potentially immutable, and will lead to opportunity capture where the lowest tax rates, the best education, and the best healthcare are claimed by the children of the rich. This creates dynamic and mutually reinforcing cycles of advantage that are transmitted across generations, making process of social mobility even harder.

Whereas high or extreme income inequality is generally perceived to have adverse effects on a society as a whole, it should be noted that low income inequality or income equality, in principle, could cause economic, social, and political problems as well. Regardless of political regime a country chooses to adopt, if everyone's income is equal or slightly different, there should be no incentives for people to be creative or try to do things differently because no matter how hard they try or what they do and/or invent, there will be no extra benefits. A hypothetical example would be to imagine that a brain surgeon doctor has the same monthly salary as a garbage collector. In such a society, it is likely that there would be labour shirking and/or free-riding problems. The social and economic consequences would be poor discipline and low initiatives among workers, poor quality and limited selection of goods and services, as well as slow technological progress (Soubbotina & Sheram, 2000). These eventually could put the whole country into an incentive trap which has negative impacts on productivity and economic growth. In addition, for socialist, autocratic, or nondemocratic countries, the time and monetary costs of top-down monitoring and enforcement should be extremely high in order to ensure that everybody has equal income or gets the same ration.[5] Except for the ruler or head of state, in a society where people are forced to have the same wage or ration, it is usually coupled with social and economic problems that could give rise to protests, riots, and/or political up-risings.

Based on the potential harmful effects of both high and low income inequality on societies as discussed above, it follows that a country where income inequality is too high should lower her income inequality while a country that has too low income inequality should increase her income inequality in order to avoid such negative effects. Viewed this way, the logic of an appropriate degree of income inequality for a country could be established. The next task is to find empirically an appropriate level of income inequality for a country. This study hypothesizes that the appropriate level of income inequality as measured by Gini coefficient for a country should be positively correlated with population size of that country. That is a country with small populations should have relatively lower Gini coefficient than a country with large populations. This is because it does not matter whether a country is underdeveloped, developing, or developed, if a country were to have only one population, income inequality of that country as measured by Gini coefficient would be zero. If a country were to have population larger than one, the chance, that income inequality as measured by Gini coefficient should rise, becomes higher due to population heterogeneity.

---

[5] This excludes the ruler or head of state who typically has extremely much larger share of income.

Having established the logic of the appropriate degree of income inequality and setting up hypothesis regarding the positive correlation between the degree of income inequality and population size, the next section explains research methodology employed to test hypothesis whether there is such a correlation. If so, what does the linkage imply about the appropriate level of income inequality of a country?

## 3. Research Methodology

The degree of heterogeneity in social, economic, and political factors, that could result in different income inequality across countries, makes it difficult to find a common set of variables that have similar effects for all countries. It is hard to argue that Singapore with population of 5.53 million should have the same social, economic, and political factors affecting income inequality as those of China

31

with population of 1.37 billion.[6] In addition, the number of those factors may not be equal for both countries at a given period of time. For these reasons, this study postulates that the degree of social, economic, and political diversities for any country could be reflected by population heterogeneity in that country. In other words, the information regarding social, economic, and political factors of a given country is already compressed in the data on the number of population of that country. This would allow us to examine the relationship between the degree of income inequality as measured by Gini coefficient and the size of population by employing ordinary least squares regression,[7] and to find out empirically the level of income inequality as measured by Gini coefficient that is appropriate for the size of population. To examine such a relationship, this study employs income inequality and population data of sixty-nine countries in the year 2012 from the World Bank (2016a; 2016b).

The following section reports the empirical evidence of the relationship between the level of income inequality as measured by Gini coefficient and population size and discusses the issue of an appropriate value of Gini coefficient for a country given size of population.

---

[6] The data on populations of Singapore and China come from the World Bank (2016b).

[7] It should be noted that the purpose of employing regression analysis is to examine the correlation between income inequality as measured by Gini coefficient and population size, not their causation. As argued in Taleb (2012), in a complex and multidimensional world, 'the notion of "cause" itself is suspect; it is either nearly impossible to detect or not really defined.'

## 4. Empirical Results

Figure 1 illustrates scatter plots of the relationship between levels of income inequality as measured by Gini coefficient and natural logarithm of population size by taking into account the possibility that if a country has only one population, then the Gini coefficient must be zero.[8] The scatter plots indicate that the relationship between the two variables should be positive.

By employing curve fitting technique, this study finds that the relationship between the level of income inequality as measured by Gini coefficient and natural logarithm of population size is nonlinear that can be best described by a second-degree polynomial function.[9] The following nonlinear equation is therefore employed to estimate the relationship between Gini coefficient and natural logarithm of population size.

$$\text{Gini} = \alpha + \beta_1 * \ln(\text{Pop}) + \beta_2 * [\ln(\text{Pop})]^2 + \varepsilon \qquad (1)$$

Where

$\alpha = 0$, $\beta_1 > 0$, $\beta_2 < 0$ and
Gini = Gini Coefficient
ln(Pop) = Natural Logarithm of Population Size
$\varepsilon$ = Error Term

---

[8] A country with one population is included in the sample in order to make the case more physically and mathematically realistic in the sense that if there were such a country, then income inequality as measured by Gini coefficient would have to be zero. This is always true regardless of social, economic, and political background of that country.

[9] The authors also tried a linear function, a third-degree polynomial function, and a nonlinear function that includes natural logarithm of population size and square root of natural logarithm of population size to fit the data points but found that a polynomial function of degree two yields the best fit. These results are available upon request.

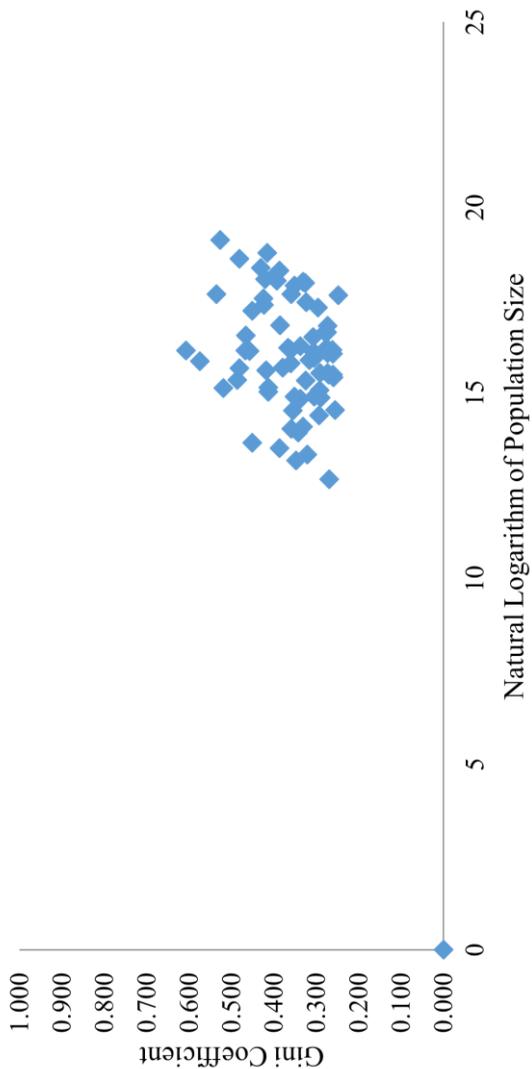

Figure 1. The Relationship between Levels of Income Inequality as Measured by Gini Coefficient and Natural Logarithm of Population Size

By using ordinary least squares estimator with heteroskedasticity-consistent standard errors and covariance, the estimated nonlinear relationship between Gini coefficient and natural logarithm of population size is as follows:

$$\text{Gini} = 0.0304*\ln(\text{Pop}) - 0.0005*[\ln(\text{Pop})]^2 + \varepsilon \qquad (2)$$

The estimated results from equation (2) and from Table 1 below indicate that the coefficient on natural logarithm of population size is statistically significant at 5 percent level while that on natural logarithm of population size square is statistically insignificant. The results confirm the hypothesis of positive correlation between level of income inequality as measured by Gini coefficient and natural logarithm of population size. A*djusted $R^2$* indicates that variations of natural logarithm of population size and of natural logarithm of population size square could explain variation of Gini coefficient around 25 percent.

**Table 1. Estimated Nonlinear Relationship between Gini Coefficient and Natural Logarithm of Population Size**

| *Explanatory Variable* | *Coefficient* | *P-Value* |
|---|---|---|
| Constant | 0 | N/A |
| Natural Logarithm of Population Size | 0.0304 (5.26) | 0.0000 |
| [Natural Logarithm of Population Size]$^2$ | -0.0005 (-1.36) | 0.1785 |

Notes: *Adjusted $R^2$* = 0.2455; *t-statistics* are in parentheses; number of countries = 69; total of observations = 70 including an

additional sample where a country has one population which would result in the Gini coefficient to be zero.

Figure 2 and Figure 3 illustrate the scatter plots between the estimated Gini coefficient and natural logarithm of population size and between the estimated Gini coefficient and natural logarithm of population size square respectively. In addition, the population sizes, the levels of actual income inequality as measured by Gini coefficient, and appropriate values of Gini coefficient estimated by this study for sixty-nine countries in 2012 are shown in Table 2.

The empirical results from Table 2 show that there are twelve out of sixty-nine countries that have the difference between the estimated Gini coefficient and the actual Gini coefficient by less than five percent.[10] If the difference between the estimated Gini coefficient and the actual Gini coefficient is allowed to be less than ten percent, there are twenty-three countries in this sample.[11] This indicates that, given countries' population sizes, about one-fifth to one-third of countries in the sample have Gini coefficients close to their appropriate values while the other two-third to four-fifth have either too high or too low Gini coefficients. As explained in Section 2, too high or too low income inequality could cause economic, social, and/or political difficulties in the society. Therefore, countries that have high income inequality should make an effort to reduce it whereas those with low income inequality should try to increase it.

---

[10] Those countries are Bulgaria, Cyprus, Estonia, Greece, Sri Lanka, Lithuania, Montenegro, Mongolia, Portugal, Thailand, Turkey, and Vietnam.

[11] In addition to twelve countries listed in the previous footnote, there are eleven more countries which are Spain, Guinea, Ireland, Italy, Lao PDR, Luxembourg, Latvia, Mauritius, Philippines, Russian Federation, and Democratic Republic of Congo.

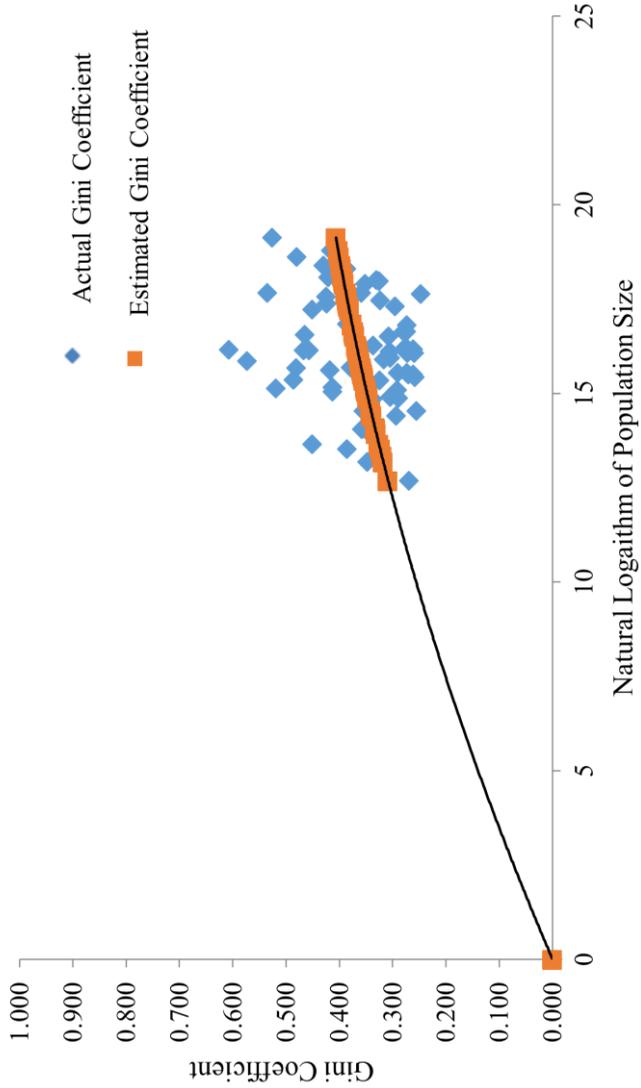

Figure 2. The Relationship between Estimated Gini Coefficient and Natural Logarithm of Population Size

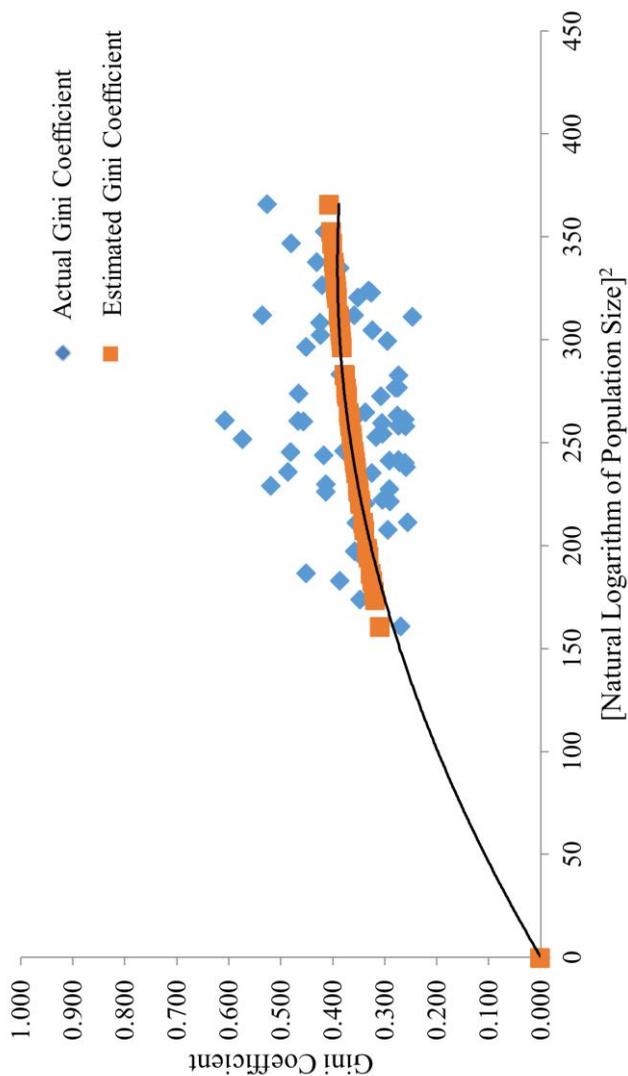

**Figure 3. The Relationship between Estimated Gini Coefficient and [Natural Logarithm of Population Size]$^2$**

Table 2. Estimated Gini Coefficient for a Given Size of Population

| Country Name | Population Size in 2012 (1) | Actual Gini in 2012 (2) | Estimated Gini (3) | (3) – (2) (Percent) |
|---|---|---|---|---|
| Albania | 2,900,489 | 0.290 | 0.342 | 17.97 |
| Argentina | 42,095,224 | 0.425 | 0.380 | -10.66 |
| Armenia | 2,978,339 | 0.305 | 0.342 | 12.22 |
| Austria | 8,429,991 | 0.305 | 0.358 | 17.34 |
| Belgium | 11,128,246 | 0.276 | 0.362 | 31.07 |
| Bulgaria | 7,305,888 | 0.360 | 0.356 | -1.26 |
| Belarus | 9,464,000 | 0.260 | 0.359 | 38.14 |
| Bolivia | 10,238,762 | 0.467 | 0.360 | -22.82 |
| Brazil | 202,401,584 | 0.527 | 0.399 | -24.34 |
| Bhutan | 743,711 | 0.387 | 0.320 | -17.31 |
| Switzerland | 7,996,861 | 0.316 | 0.357 | 12.79 |
| Colombia | 46,881,018 | 0.535 | 0.381 | -28.84 |
| Costa Rica | 4,654,148 | 0.486 | 0.349 | -28.23 |
| Cyprus | 1,129,303 | 0.343 | 0.327 | -4.82 |
| Czech Republic | 10,510,785 | 0.261 | 0.361 | 38.08 |
| Djibouti | 853,069 | 0.451 | 0.322 | -28.67 |
| Denmark | 5,591,572 | 0.291 | 0.352 | 20.92 |
| Dominican Republic | 10,155,036 | 0.457 | 0.360 | -21.12 |

Table 2. (Continued)

| Country Name | Population Size in 2012 (1) | Actual Gini in 2012 (2) | Estimated Gini (3) | (3) – (2) (Percent) |
|---|---|---|---|---|
| Ecuador | 15,419,493 | 0.466 | 0.366 | -21.37 |
| Spain | 46,773,055 | 0.359 | 0.381 | 6.14 |
| Estonia | 1,322,696 | 0.332 | 0.329 | -0.71 |
| Finland | 5,413,971 | 0.271 | 0.351 | 29.48 |
| France | 65,639,975 | 0.331 | 0.385 | 16.37 |
| United Kingdom | 63,700,300 | 0.326 | 0.385 | 18.15 |
| Georgia | 3,825,000 | 0.414 | 0.346 | -16.35 |
| Guinea | 11,628,767 | 0.337 | 0.362 | 7.39 |
| Greece | 11,045,011 | 0.367 | 0.362 | -1.44 |
| Honduras | 7,736,131 | 0.574 | 0.356 | -37.91 |
| Haiti | 10,288,828 | 0.608 | 0.360 | -40.70 |
| Hungary | 9,920,362 | 0.306 | 0.360 | 17.83 |
| Ireland | 4,586,897 | 0.325 | 0.349 | 7.21 |
| Iraq | 32,780,975 | 0.295 | 0.376 | 27.40 |
| Iceland | 320,716 | 0.269 | 0.305 | 13.23 |
| Italy | 59,539,717 | 0.352 | 0.384 | 9.21 |
| Kazakhstan | 16,791,425 | 0.274 | 0.367 | 33.93 |
| Kyrgyz Republic | 5,607,200 | 0.274 | 0.352 | 28.49 |

Table 2. (Continued)

| Country Name | Population Size in 2012 (1) | Actual Gini in 2012 (2) | Estimated Gini (3) | (3) − (2) (Percent) |
|---|---|---|---|---|
| Cambodia | 14,832,255 | 0.308 | 0.366 | 18.87 |
| Kosovo | 1,805,200 | 0.294 | 0.334 | 13.67 |
| Lao PDR | 6,473,050 | 0.379 | 0.354 | -6.63 |
| Sri Lanka | 20,424,000 | 0.386 | 0.370 | -4.09 |
| Lithuania | 2,987,773 | 0.352 | 0.342 | -2.67 |
| Luxembourg | 530,946 | 0.348 | 0.314 | -9.79 |
| Latvia | 2,034,319 | 0.355 | 0.336 | -5.28 |
| Moldova | 3,559,519 | 0.292 | 0.345 | 18.25 |
| Mexico | 122,070,963 | 0.481 | 0.393 | -18.31 |
| Montenegro | 620,601 | 0.322 | 0.317 | -1.58 |
| Mongolia | 2,808,339 | 0.338 | 0.341 | 1.08 |
| Mauritius | 1,255,882 | 0.358 | 0.328 | -8.40 |
| Netherlands | 16,754,962 | 0.280 | 0.367 | 31.24 |
| Norway | 5,018,573 | 0.259 | 0.350 | 35.14 |
| Panama | 3,743,761 | 0.519 | 0.346 | -33.41 |
| Peru | 30,158,768 | 0.451 | 0.375 | -16.81 |
| Philippines | 96,017,322 | 0.430 | 0.390 | -9.42 |
| Poland | 38,063,164 | 0.324 | 0.378 | 16.79 |

Table 2. (Continued)

| Country Name | Population Size in 2012 (1) | Actual Gini in 2012 (2) | Estimated Gini (3) | (3) – (2) (Percent) |
|---|---|---|---|---|
| Portugal | 10,514,844 | 0.360 | 0.361 | 0.11 |
| Paraguay | 6,379,162 | 0.482 | 0.354 | -26.60 |
| Romania | 20,058,035 | 0.273 | 0.370 | 35.31 |
| Russian Federation | 143,201,676 | 0.416 | 0.395 | -5.13 |
| El Salvador | 6,072,233 | 0.418 | 0.353 | -15.59 |
| Slovak Republic | 5,407,579 | 0.261 | 0.351 | 34.43 |
| Slovenia | 2,057,159 | 0.256 | 0.336 | 31.40 |
| Sweden | 9,519,374 | 0.273 | 0.359 | 31.55 |
| Thailand | 67,164,130 | 0.393 | 0.385 | -1.81 |
| Turkey | 74,099,255 | 0.402 | 0.387 | -3.74 |
| Uganda | 35,400,620 | 0.424 | 0.377 | -10.94 |
| Ukraine | 45,593,300 | 0.247 | 0.381 | 53.84 |
| Uruguay | 3,396,753 | 0.413 | 0.344 | -16.73 |
| Vietnam | 88,809,200 | 0.387 | 0.389 | 0.49 |
| Congo, Dem. Rep. | 70,291,160 | 0.421 | 0.386 | -8.30 |
| Hypothetical Country | 1 | 0.000 | 0.000 | 0.00 |

Sources: The World Bank (2016a; 2016b) and the authors' calculation.

However, this does not mean that countries that have the levels of income inequality as measured by Gini coefficient equal or close to the appropriate levels should stay passive. It is possible that, given approximately equal sizes of population and Gini coefficients, the ratio of income share held by the rich to the income share held by the poor in one country is much higher than that of the other country. In this case, the former country should come up with public policies in order to reallocate income among populations by increasing income of the poor and at the same time reducing income of the rich in such a way that the targeted or appropriate level of income inequality remains unchanged.

## 5. Conclusions, Policy Implications, and Suggestions for Future Research

This study views that the pursuit of having an appropriate level of income inequality should be considered as one of the biggest challenges facing academic scholars as well as policy makers. Unfortunately, technical and empirical research on this particular issue are currently lacking. As a result, most, if not all, policy attempts by governments around the world to either reduce or raise the level of income inequality (mostly reducing) are designed and implemented without prior knowledge about targeted Gini coefficients in mind. By employing the World Bank's data on population size to reflect the heterogeneity in economic, social, and political factors as well as to level playing field among countries and on Gini coefficient, the logic and empirical findings of appropriate levels of income inequality as measured by Gini coefficient for sixty-nine countries from this study could be used as a guideline for policymakers before designing and conducting public policies in order to pursue the targeted level of income inequality. This study conjectures that, for a given population size, countries that achieve the targeted level of income inequality should perform better in terms of economic growth and well-being than those that are far away from their appropriate levels of income inequality.

In addition, the issue of widening gap between income (and/or wealth) share held by the rich and income (and/or wealth) share held by the average population has recently caught public attention. For example, according to Frank (2011), heads of the largest corporations in the United States of America at present earn four hundred times as much as average workers, compared to forty times as much back in 1980s. Research conducted by Oxfam also indicates that, in 2015, the richest sixty-two people in the world own half of global wealth (Reuben, 2016). While the main focus among academic scholars and policymakers has been on the issue of how to narrow the income (and/or wealth) gap between the richest and the poorest, this study believes otherwise. It hypothesizes that the root of the problem may not lie between income (and/or wealth) gap of the richest and the poorest, but rather that of the richest and the second richest. The theoretical idea behind this is that when income (and/or wealth) of the richest group gets larger than that of the second richest group up to the point that passes a critical threshold, it could make the second richest group feels that it is unfair. It might also be possible that the richest group feels that their economic, social, and/or political statuses are threatened by the second richest group. The battle between the two hegemonic groups could cause chaos in the society mainly because both the richest and the second richest have all the resources to influence government policies, to bend laws, regulations, and constitutions, to create large-scale propaganda and conflicts of memes among interest groups and grassroot people, as well as to shape referendum or election outcome. It is of interest to examine whether this hypothesis is rejected or not. If not, then it is worth to find out what an appropriate income (and/or wealth) gap between the richest and the second richest groups that yields no conflict between these two hegemonic groups for the good of the society.

Moreover, it is of challenge to search for appropriate gaps of percentage share of income among subgroups of population in the society such that those who have lower income feel wholeheartedly that it is fair and square for them to have less income than those who earn more. These interesting issues await future research.[12]

---

[12] The authors thank Suradit Holasut for pointing out these issues.

# References

Alesina, A. (2003). The Size of Countries: Does It Matter?. *The Journal of the European Economic Association*, 1 (2-3), pp. 301-316.

Bräutigam, D. & Woolcock, M. (2001). Small States in a Global Economy: The Role of Institutions in Managing Vulnerability and Opportunity in Small Developing Countries. *WIDER Discussion Paper No. 2001/37*. Retrieved from http://www.wider.unu.edu/sites/default/files/dp2001-37.pdf

Campante, F. & Do, Q. A. (2007). Inequality, Redistribution, and Population. *KSG Faculty Research Working Paper Series No. RWP07-046*. Retrieved from https://research.hks.harvard.edu/publications/workingpapers/citation.aspx?PubId=4995&type=WPN

Commonwealth Secretariat. (2000). Small States: Meeting Challenges in the Global Economy. *Report of the Commonwealth Secretariat/World Bank Joint Task Force on Small States*. Retrieved from http://documents.worldbank.org/curated/en/267231468763824990/Small-states-meeting-challenges-in-the-global-economy

Deltas, G. (2003). The Small-Sample Bias of the Gini Coefficient: Results and Implications for Empirical Research. *The Review of Economics and Statistics*, 85 (1), pp. 226-234.

Frank, R. H. (2011). *The Darwin Economy: Liberty, Competition, and the Common Good*. New Jersey: Princeton University Press.

Fuentes-Nieva, R. & Galasso, N. (2014). Working for the Few: Political Capture and Economic Inequality. *Oxfam Briefing Paper*. Retrieved from https://www.oxfam.org/en/research/working-few

Gini Coefficient. (n.d.). In *Wikipedia*. Retrieved July 7, 2016, from https://en.wikipedia.org/wiki/Gini_coefficient

Phongpaichit, P. (2016). Commentary Note on Thailand's Current Inequality Situation and Its Prospects. *Thammasat Review of Economic and Social Policy*, 2 (1), pp. 4-16.

Raza, S. A. (2016). r > g: Increasing Inequality of Wealth and Income Is a Runaway Process. In *2016: What Do You Consider The Most Interesting Recent [Scientific] News? What Makes It Important?*. Retrieved from https://www.edge.org/annual-questions

Reuben, A. (2016, January 16). Wealth of Richest 1% 'Equal to Other 99%'. *BBC News*. Retrieved from http://www.bbc.com/news/business-35339475

Soubbotina, T. P. & Sheram, K. A. (2000). *Beyond Economic Growth: Meeting the Challenges of Global Development*. Washington, D.C.: The World Bank.

Streeten, P. (1993). The Special Problems of Small Countries. *World Development*, 21 (2), pp. 197-202.

Taleb, N. N. (2012). *Antifragile: Things That Gain from Disorder*. New York: Random House.

The World Bank. (2016a). *Gini Index (World Bank Estimate)* [Data file]. Retrieved from http://data.worldbank.org/indicator/SI.POV.GINI

The World Bank. (2016b). *Population Ranking* [Data file]. Retrieved from http://databank.worldbank.org/data/reports.aspx?Code=SP.POP.TOTL&id=af3ce82b&report_name=Popular_indicators&populartype=series&ispopular=y